# High $T_c$ State of Lanthanum Hydrides

Hao Song[1], Defang Duan[1,*], Tian Cui[1,2], Vladimir Z. Kresin[3,*]

[1]*State Key Laboratory of Superhard Materials, College of Physics,*

*Jilin University, Changchun 130012, China*

[2]*School of Physical Science and Technology, Ningbo University, Ningbo, 315211, China*

[3]*Lawrence Berkeley Laboratory, University of California at Berkeley,*

*Berkeley, CA 94720, USA*

*Corresponding authors' email: duandf@jlu.edu.cn, vzkresin@lbl.gov

**Abstract:**

The high $T_c$ phase of lanthanum hydrides deserves special attention because this phase displays the highest observed critical temperature. We focus on the evaluation of the critical temperature, isotope coefficient and their pressure dependence of this phase ($Fm\bar{3}m$ phase of the $\text{LaH}_{10}$ compound). By combining the method of two coupling constants, $\lambda_{\text{opt}}$ and $\lambda_{\text{ac}}$, and two characteristic frequencies with first-principle calculations we are able to analyze the role of the light hydrogen in the amplitude of the critical temperature in this unique compound. Specifically, the critical temperature of 254 K and the isotope coefficient of 0.45 are evaluated at high pressure, and the results are in good agreement with the experimental data. The peculiar decrease in $T_c$ upon further increase in pressure is explained by the presence of a flat region on the Fermi surface and the appearance of a two-gap structure, which should be experimentally observable by tunneling spectroscopy.

**I. INTRODUCTION**

This paper is concerned with properties of the high $T_c$ superconducting hydrides. Our study was motivated by the recent observation [1,2] of the record value of $T_c \simeq 250 - 260$ K (!) for the $LaH_{10}$ compound. This progress with the La – based hydrides represents further development following the discovery of high $T_c$ sulfur hydride with its $T_c \simeq 203$ K [3]. Note that the situation with new high $T_c$ family, hydrides, is entirely different from that with other well known class of high $T_c$ materials, cuprates. Indeed, the discovery of the high $T_c$ superconductivity in the cuprates [4] was totally unexpected. Contrary to this, the experimental observation [3] was preceded by theoretical investigation and key predictions (see review [5]). All predictions were based on the phonon mechanism, which was generally accepted shortly after the experimental observations of the transition into superconducting state and also large values of the isotope coefficient. On the other site, the mechanism of superconductivity in the cuprates is still a controversial issue.

The initial idea [6] that the metallic hydrogen (the transition into metallic state requires high pressure) can display the high $T_c$ superconducting state was generalized for the compounds containing hydrogen [7]. A number of interesting theoretical and experimental studies were performed in the early years of this century. For example, the first experimental studies of $SiH_4$ demonstrated $T_c \simeq 17.5$ K at $P \simeq 96$ GPa [8,9]. Later the theoretical studies [10,11] predicted for $(SiH_4)(H_2)_2$ and $CaH_6$ the values of $T_c \simeq 107$ K and $T_c \simeq 235$ K, respectively.

Speaking of sulfur hydrides, it was predicted [12] that increase in pressure leads to formation of the $H_3S$ compound and it is accompanied by large increase in a number of hydrogen ligands in the unit cell. As a result, one can observe the value of $T_c \simeq 200$ K. The experiments [3] totally confirmed this prediction, so that the record high value of $T_c$ was observed. The initial increase up to $T_c \simeq 90$ K observed in [3] was in agreement with the theoretical calculations [13].

The observation of the high $T_c$ superconducting state in the $H_3S$ hydride was followed by intensive theoretical development. Since the pairing is caused by the electron-phonon interaction, the analysis is based on the Eliashberg equation [14]. As we know, this equation does not contain explicitly the coupling constant $\lambda$. Correspondingly, the value of the critical temperature can be found without invoking the coupling constant concept [15-17]. The evaluation was performed using the superconducting density functional theory [18,19]. The study of multi-phase compound was described in [20]. This study is very important, since the pressure leads to sequence of structural transitions, and the situation when the compound is characterized by the presence of multi-phase structure is rather typical.

The family of hydrides is rather large and it was generally expected that further studies will result in an observation of even higher values of $T_c$. The recent experimental observation of record high $T_c$ in lanthanum hydride [1,2] was preceded by theoretical studies [21,22]. As for the experimental discovery, it is essential that the paper [2] contains not only the description of the drastic change in resistance and the impact of the magnetic field on the value of the energy gap, but also the observation of the isotope effect.

The observation of high $T_c$ in the $H_3S$ compound was followed by an intensive theoretical development. Similarly, a detail study of the $LaH_{10}$ compound is of definite interest. The paper [23] is concerned with an anharmonicity and its impact on $T_c$. Note that this interesting problem was analyzed earlier in [24-28], see also the review [29], Sec.3.9. However, the approach used in [23] appears to be inconsistent and this will be discussed in detail below, Sec. IV.

As was noted above, the present paper is motivated by recent observation of superconductivity in lanthanum hydride and is concerned with fundamentals of this new compound with record high $T_c$. More specifically, we focus on the evaluation of the critical temperature and its peculiar pressure dependence in the high $T_c$ phase. We calculate also the isotope coefficient at various pressures; its value is directly related to the interplay of the optical and acoustic phonon modes.

An analysis contains *ab initio* calculations similar to those used by two of us with our collaborators in [5,12]. It is based also on the two-coupling constant method developed by Gor'kov and one of the co-authors [30,31]. Our goal is to develop the self-consistent description of the high $T_c$ phase of the $\text{LaH}_{10}$ compound and to determine some features which are common for the new family of the high $T_c$ hydrides.

## II. CALCULATION DETAILS

Structure relaxations, electronic properties, elastic constants are calculated using the VASP package [32] in the framework of DFT with Perdew–Burke–Ernzerhof parameterization of the generalized gradient approximation [33]. Elastic constants were evaluated using linear relationship between stress and strain. We employed projector augmented waves (PAW) method to model the interactions between electrons and ions, and H $1s^1$ and La $5s^25p^65d^16s^2$ were regarded as valence electrons. We applied a kinetic energy cutoff of 800 eV and a Brillouin zone sampling grid spacing of 2π×0.03 Å$^{-1}$ for structure relaxations and 2π×0.02 Å$^{-1}$ for electronic property calculations. Lattice dynamics and electron–phonon coupling (EPC) were characterized by density functional perturbation theory as implemented in the Quantum-ESPRESSO package [34]. Norm-conserving potentials for H ($1s^1$) and La ($5s^2 5p^6 5d^1 6s^2$) were used with a kinetic energy cutoff of 80 Ry. In the calculations of EPC parameter $\lambda$, the first Brillouin zone was sampled using a 6×6×6 q-points mesh with a denser 24×24×24 k-points mesh (with Gaussian smearing and σ = 0.035 Ry which approximates the zero-width limits in the calculation of $\lambda$). These above parameters settings were tested strictly to ensure the $T_c$s converged to less than 1 K.

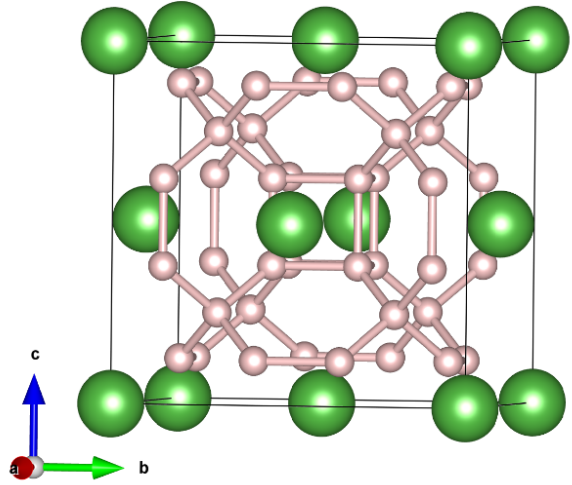

FIG. 1. (Color online) Structure of the high $T_c$ $\text{LaH}_{10}$ compound ($Fm\bar{3}m$). La atoms are colored green, and H atoms are pink.

### III. RESULTS AND DISSCUSSION

### III. A. Main equations and critical temperature

Pressure is a thermodynamic parameter, and, as we know, its increase leads initially to formation of the metallic state and then, at higher pressures, to sequence of structural transitions. We will focus on the $Fm\bar{3}m$ phase of the $\text{LaH}_{10}$ compound with its record high $T_c$ value (Fig. 1). In this structure, La atoms occupy the 4a (0.0 0.0 0.0) sites, while H atoms occupy 8c (0.25, 0.25, 0.25) and 32f (0.12, 0.12, 0.38) sites. The structure relaxations at 200, 250, and 300 GPa yield the lattice parameter $a$ = 4.93, 4.84, and 4.75 Å, respectively.

This choice of the phase is not occidental. Indeed, the calculation of enthalpy shows that the phase diagram is a complex one. At P<200 GPa, we are dealing with the competition of the $R\bar{3}m$ and $C2/m$ phases, as shown in Fig. 2. In this region of pressure the system is described by the percolation picture (see, e.g., [20,35]). As for the pressures P>200 GPa, the system is in the $Fm\bar{3}m$ phase, which can be studied by analogy with the $Im\bar{3}m$ phase of the sulfur hydrides. Moreover, exactly this phase displays the highest observed value of $T_c$.

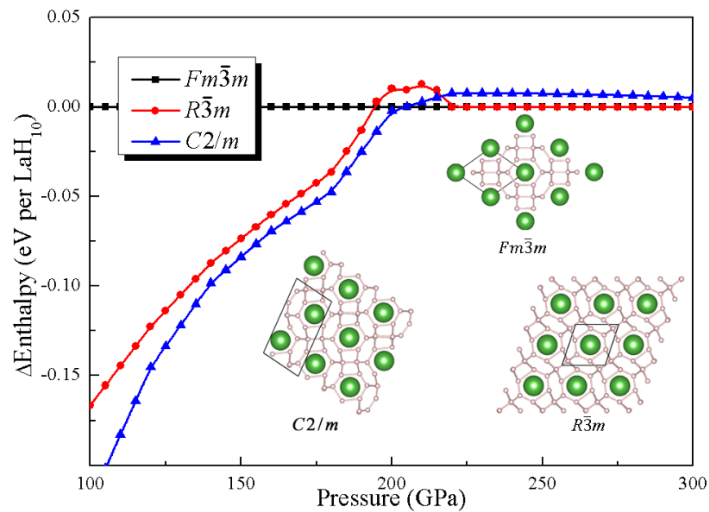

FIG. 2. (Color online) Calculated enthalpy curves for $C2/m$ and $R\bar{3}m$ relative to $Fm\bar{3}m$ as a function of pressure. They are competitive phases which have lower enthalpy value than $Fm\bar{3}m$ below 200 GPa.

As a starting point, we are using the main equation allowing to evaluate the value of the critical temperature, if the superconducting state is provided by the electron-phonon interaction. An excellent agreement between the theoretical predictions and experimental data on $H_3S$ system as well as the observation of a large isotope effect for the $H \to D$ substitution [2] provide strong evidence for the electron-phonon interaction being the mechanism of superconductivity in hydrides. The equation has a form (at $T = T_c$) [14]:

$$\Delta(\omega_n)Z = \pi T_c \sum_m d\Omega \frac{\alpha^2(\Omega)F(\Omega)}{\Omega} D(\Omega, \omega_n - \omega_m) \frac{\Delta(\omega_m)}{|\omega_m|}\bigg|_{T_c} \quad (1)$$

$$Z = 1 + (\pi T_c/\omega_n) \sum_m \int d\Omega \frac{\alpha^2(\Omega)F(\Omega)}{\Omega} D(\Omega, \omega_n - \omega_m) \frac{\omega_m}{|\omega_m|}\bigg|_{T_c} \quad (2)$$

Here $\Delta(\omega_n)$ is the thermodynamic order parameter (we employ the method of the thermodynamic Green's functions (see, e.g., [36,37]); $\omega_n = (2n + 1)\pi T_c$, $Z$ is the renormalization function, and $D(\Omega)$ is the phonon propagator.

$$D(\Omega) = \frac{\Omega^2}{\Omega^2 + (\omega_n - \omega_{n\prime})^2} \quad (3)$$

The eqs. (1) and (2) contain the function $g(\Omega) = \alpha^2(\Omega)F(\Omega)$ and Coulomb pseudopotential $\mu^*$; $F(\Omega)$ is the phonon density of states, and $\alpha^2(\Omega)$ contains the electron-phonon matrix element (see, e.g., [38,39]).

In order to obtain an analytical expression for $T_c$, one should introduce the coupling constant $\lambda$, which is not explicitly enters Eq.(1). Usually this quantity can be introduced, because the phonon propagator $D(\Omega, \omega_n - \omega_{n\prime})$ is a smooth function relative to the phonon density of states, $F(\Omega)$. It allows to replace $D(\Omega, \omega_n - \omega_{n\prime}) \simeq D(\widetilde{\Omega}, \omega_n - \omega_{n\prime})$, $\widetilde{\Omega}$ is the average phonon frequency (see [40-42]), and introduce the coupling constant, defined by the relation

$$\lambda = 2 \int d\Omega \frac{\alpha^2(\Omega)F(\Omega)}{\Omega}; \quad (4)$$

The average phonon frequency $\widetilde{\Omega}$ is defined by the expression:

$$\widetilde{\Omega} = \langle \Omega^2 \rangle^{1/2}; \quad \langle \Omega^2 \rangle = \frac{2}{\lambda} \int d\Omega\, \Omega \alpha^2(\Omega)F(\Omega); \quad (4')$$

One can select also the close value $\widetilde{\Omega}_{\log}$, see [43,44]. Below we will be using the value $\widetilde{\Omega}$, defined by eq. (4').

The hydrides are characterized by complex phonon spectrum; it contains acoustic and optical modes. Because of the presence of light hydrogen ions, we are dealing with high frequency optical modes. As a result, the phonon spectrum is very broad. As a consequence, it is more accurate to introduce two characteristic frequencies $\widetilde{\Omega}_{ac}$ and $\widetilde{\Omega}_{opt}$ and two coupling constants $\lambda_{opt}$ and $\lambda_{ac}$ describing the interaction of electrons with optical and acoustic phonons, respectively [30,31].

Let us select the value of the characteristic phonon frequency $\Omega_1$, which separates optical and acoustic modes. As one can see from Table I, its value is different at various pressures. The equation (1) can be written in the form:

$$\Delta(\omega_n)Z = \pi T \sum_{\omega_{n'}} \left[ \lambda_{opt} \frac{\widetilde{\Omega}_{opt}^2}{\widetilde{\Omega}_{opt}^2 + (\omega_n - \omega_{n'})^2} \right.$$

$$\left. + \lambda_{ac} \frac{\widetilde{\Omega}_{ac}^2}{\widetilde{\Omega}_{ac}^2 + (\omega_n - \omega_{n'})^2} \right] \frac{\Delta(\omega_{n'})}{|\omega_{n'}|} \Bigg|_{T=T_c} \quad (1')$$

and corresponding equation for the renormalization factor $Z$.
Here

$$\lambda_{ac} = 2 \int_0^{\Omega_1} d\Omega \frac{\alpha^2(\Omega)F(\Omega)}{\Omega} \quad (5)$$

$$\lambda_{opt} = 2 \int_{\Omega_1}^{\Omega_{max}} d\Omega \frac{\alpha^2(\Omega)F(\Omega)}{\Omega} \quad (6)$$

The values of $\lambda_{ac}$ and $\lambda_{opt}$ can be determined directly from the dependence $\lambda(\Omega)$ (see, e.g., Fig. 5 in [12], and below, Fig. 3). More specifically, $\lambda_{ac} = \lambda(\Omega_1)$, and $\lambda_{opt} = \lambda_{max} - \lambda_{ac}$.

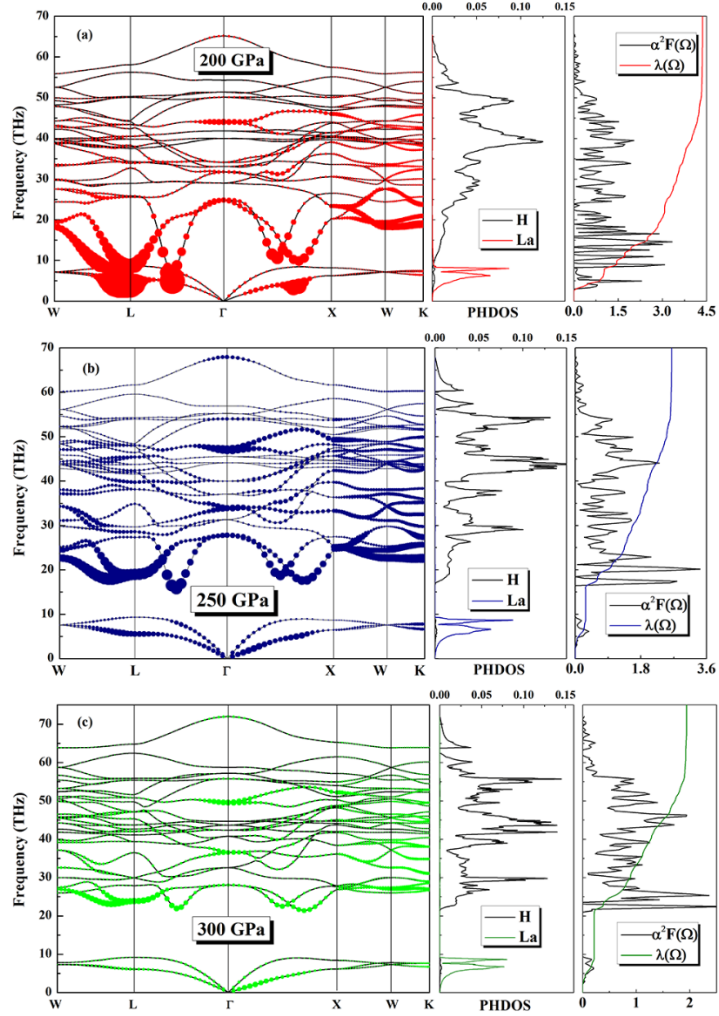

FIG. 3. (Color online) Phonon dispersion, phonon density of states (PDOS), spectral function $\alpha^2(\Omega)F(\Omega)$ and the electron-phonon integral $\lambda(\Omega)$ in $Fm\bar{3}m$ phase of LaH$_{10}$ at a) P= 200 GPa; b) P = 250 GPa; and c) P = 300 GPa. The color solid circles projected on phonon bands show mode-resolved electron-phonon coupling parameters $\lambda_{q,\nu}$ and the size of the circle is proportional to EPC strength.

As for the quantities $\widetilde{\Omega}_{ac}$ and $\widetilde{\Omega}_{opt}$, they are defined by the following relation:

$$\widetilde{\Omega}_{ac} = \langle \Omega_{ac}^2 \rangle^{1/2}; \langle \Omega_{ac}^2 \rangle = \frac{2}{\lambda_{ac}} \int_0^{\Omega_1} d\Omega \cdot \Omega^2 \frac{\alpha^2(\Omega)F(\Omega)}{\Omega}; \qquad (7)$$

$$\widetilde{\Omega}_{opt} = \langle \Omega_{opt}^2 \rangle^{1/2}; \langle \Omega_{opt}^2 \rangle = \frac{2}{\lambda_{opt}} \int_{\Omega_1}^{\Omega_{max}} d\Omega \cdot \Omega^2 \frac{\alpha^2(\Omega)F(\Omega)}{\Omega}; \qquad (8)$$

We will be using the average values determined by the relations (7) and (8). They are generalization of the definitions for the usual systems described by single coupling constant (see eq. (4)). The spectral functions $\alpha^2(\Omega)F(\Omega)$ are calculated for various pressures. Using these dependences, one can evaluate the values of $\widetilde{\Omega}_{ac}$ and $\widetilde{\Omega}_{opt}$.

Using the calculated parameters $\lambda_i, \widetilde{\Omega}_i$ ($i \equiv$ opt, ac) and eq. (1'), one can evaluate the value of the critical temperature. In the important case when the main contribution comes from the optical modes (then $\lambda_{opt} \gg \lambda_{ac}$), one can use the following expression for $T_c$ [30,31]:

$$T_c = \left[1 + 2\frac{\lambda_{ac}}{\lambda_{opt}-\mu^*} \cdot \frac{1}{1+\rho^{-2}}\right] T_c^0 \qquad (9)$$

Here $\rho = \widetilde{\Omega}_{ac}/\pi T_c^0$; $T_c^0$ is the value of the critical temperature determined by the electron-phonon interaction with the optical modes only.

As mentioned above, eq. (9) is valid, if $\lambda_{opt} \gg \lambda_{ac}$. This is the case for the high $T_c$ $Im\bar{3}m$ phase of the $H_3S$ compound; then, according to the paper [12] $\lambda_{opt} \simeq 1.5$, $\lambda_{ac} \simeq 0.5$. For such a value of $\lambda_{opt}$, one can use the well-known McMillan-Dynes expression for $T_c^0$ [35,45]; the value $T_c^0$ =175 K was obtained [31]. The additional contribution is provided by acoustic modes. Eq. (9) can be used in order to evaluate the value of the isotope coefficient (see below).

**B. Lanthanum hydride**

Let us focus on lanthanum hydride. Its high $T_c$ phase, $Fm\bar{3}m$ (Fig. 1) displays the highest observed value of the critical temperature. The structure $Fm\bar{3}m$ is formed as a result of the structural transition which occurs at $P \simeq 200$GPa. It is interesting that the transition occurs at pressures similar to those for the transition in $H_3S$ system into its high $T_c$ Phase $Im\bar{3}m$ with its cubic structure. However, there is a key difference between the sulfur hydrides and the $LaH_{10}$. The $Fm\bar{3}m$ phase has the clathrate structure (Fig. 1) with large number of hydrogen ligands in unit cell and larger strength of the electron-phonon interaction. Below we describe the properties of the superconducting state of this high $T_c$ phase $Fm\bar{3}m$, including its evolution with further increase in pressure. The analysis is based on the method described in the previous section.

The spectral function $\alpha^2(\Omega)F(\Omega)$ appears to be different at various pressures. As a first step, this function was calculated at $P = 200$ GPa, see Fig. 3a. As a next step, with use of eqs. (7) and (8), one can evaluate the characteristic frequencies $\widetilde{\Omega}_{ac}$ and $\widetilde{\Omega}_{opt}$ and the corresponding coupling constants $\lambda_{ac}$ and $\lambda_{opt}$. For $P = 200$ GPa they are: $\widetilde{\Omega}_{ac} = 2.2 \times 10^2$ K, $\widetilde{\Omega}_{opt} = 1.25 \times 10^3$ K; $\lambda_{ac} = 1.1$ and $\lambda_{opt} = 3.25$. The phonon density of states is shown on Fig. 3a; accordingly, one can select $\Omega_1 = 8.3$ THz.

Let us evaluate the value of the critical temperature. Since $\lambda_{opt} \gg \lambda_{ac}$, one can use eq. (9). Here $T_c^0$ is the value of the critical temperature caused by the interaction with optical modes only. The value of $\lambda_{opt}$ is rather large: $\lambda_{opt} = 3.25$; for such a large value the Allen-Dynes-McMillan (ADM) expression for $T_c^0$ is not applicable, since it is valid for $\lambda \lesssim 1.5$ only (the ADM expression was applicable for the $Im\bar{3}m$ phase of the $H_3S$ compound, see [30,31]).

One should use a different expression valid at large $\lambda_{\text{opt}}$. We use the expression [46], see the reviews [42,47]:

$$T_c^0 = \frac{0.25\widetilde{\Omega}_{\text{opt}}}{[e^{2/\lambda_{\text{eff}}} - 1]^{1/2}}; \qquad (10)$$

$$\lambda_{\text{eff}} = (\lambda_{\text{opt}} - \mu^*)\big[1 + 2\mu^* + \lambda_{\text{opt}}\mu^* t(\lambda_{\text{opt}})\big]^{-1}$$

$$t(x) = 1.5 \exp(-0.28\,x)$$

Using eqs. (9,10), we calculated the values of the critical temperatures at various pressures. Specific values of the critical temperature $T_c$ are presented in Table I. As one can see from Table I, the maximum value of $T_c$ turns out to be equal to $T_c \simeq 254$ K at P = 200 GPa. The values of $T_c^0$ which enter the expression for $T_c$ can be determined from Eq. (10): $T_c^0 \simeq 240$ K for P= 200GPa; $T_c^0 \simeq 236$ K for P = 250 GPa, and $T_c^0 \simeq 200$ K for P = 300 GPa (we assumed $\mu^* = 0.13$). The calculated values of $T_c$ depend on the parameters introduced above, and also on $\mu^*$. For $\mu^* = 0.1$ we obtain $T_c \simeq 270$ K (at P $\simeq$ 200 GPa ); $T_c \simeq 260$ K (at P $\simeq$ 250 GPa); and $T_c \simeq 230$ K (at P $\simeq$ 300 GPa). On the whole these temperatures are in very good agreement with the experimental data.

**Table I**. The coupling constants $\lambda_{\text{opt}}$ and $\lambda_{\text{ac}}$ in $Fm\bar{3}m$ LaH$_{10}$ at different pressures, and the values of the critical temperature.

| P (GPa) | $\Omega_1$ (THZ) | $\lambda_{\text{opt}}$ | $\lambda_{\text{ac}}$ | $\widetilde{\Omega}_{\text{opt}}$ (10$^3$K) | $\widetilde{\Omega}_{ac}$ (10$^2$K) | $T_c$ (K, $\mu^* = 0.13$) |
|---|---|---|---|---|---|---|
| 200 | 8.3 | 3.25 | 1.1 | 1.25 | 2.2 | 254 |
| 250 | 14.7 | 2.35 | 0.3 | 1.6 | 2.9 | 244 |
| 300 | 20 | 1.75 | 0.2 | 1.8 | 3.0 | 210 |

**C. Maximum $T_c$ and its decrease at higher pressures**

As is known, an increase in pressure results in a transition into the $Im\bar{3}m$ phase of the H$_3$S compound [12]. As observed in [48], with a further increase in pressure $T_c$ decreases.

A similar scenario occurs for the lanthanum hydrides. Raising the pressure up to P=200 GPa increases $T_c$ [1]; but raising the pressure further begins to reduce $T_c$ (see Table I), so that P ≃ 250 GPa and also at P ≃ 300 GPa it becomes lower than at P ≃ 200 GPa. ($T_c$ ≃ 254 K)

Such a temperature dependence can be understood as follows. The decrease in $T_c$ corresponds to a lower coupling constant $\lambda_{\text{opt}}$ which, in turn, is related to a lower value of the characteristic frequency $\widetilde{\Omega}_{\text{opt}}$ at P ≃ 200 GPa (see Table I). Indeed, in accordance with the McMillan relation [35], one can write:

$$\lambda_{\text{opt}} = \frac{\langle I^2 \rangle \nu}{M \widetilde{\Omega}_{\text{opt}}^2} \qquad (12)$$

One expects that a decrease in $\widetilde{\Omega}_{\text{opt}}$ leads to an increase in $\lambda_{\text{opt}}$; this is the so-called "softening mechanism" [35]. Thus a lower phonon frequency results in a higher $T_c$ (see the discussion in, e.g., [47], Sec.3.4.2).

The phonon frequency $\widetilde{\Omega}_{\text{opt}}$ can be expressed in terms of the elastic constant. The picture is self-consistent, since our calculation of the elastic constants for LaH$_{10}$ demonstrates that $C_{44}$ ≃ 310 kbar at P = 200 GPa, whereas $C_{44}^{\text{P}=250\text{ GPa}}$ = 615 kbar and $C_{44}^{\text{P}=300\text{ GPa}}$ = 668 kbar, $C_{44}^{\text{P}=300\text{ GPa}}$ = 668 kbar. Therefore, indeed, the value of $C_{44}$ at P ≃ 200 GPa is noticeably lower than that at higher pressures. This reduction in $C_{44}$ is related to features of the electronic spectrum of the material.

The electronic spectra in Fig. 4 show that there are regions on the Fermi surface containing peaks in the electronic density of the states (DOS). These peaks correspond to flat regions of the Fermi surface. The interaction of phonons with these regions leads to a softening of the phonon frequency (see, e.g., [49,50]), and eventually can even lead to an instability: a charge density wave transition upon a decrease in temperature. This transition is known to manifest itself in the opening of an energy gap near the flat region.

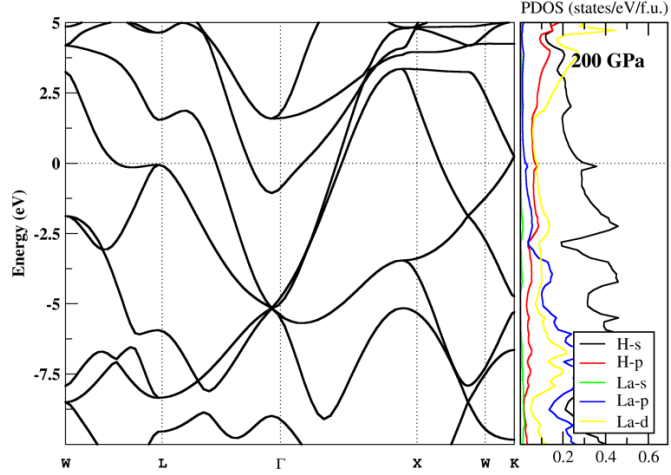

FIG. 4. (Color online) Band structure and the density of electronic states (DOS) for the high $T_c$ phase $Fm\bar{3}m$ at P = 200 GPa

In our case the pairing interaction, involving the electron-phonon interaction covering a whole wide band, dominates, so that the transition into the superconducting state occurs at $T_c$ which is above the temperature of the CDW transition. Nevertheless, the phonon frequency does decrease somewhat, as described above. The scale of this decrease depends on the intensity of the interaction with the electronic states near the flat region and, consequently, on the amplitude of the peak in the electronic DOS. The DOS calculation shows (Fig. 5) that the peak has a maximum value at P = 200 GPa and decreases at large pressures.

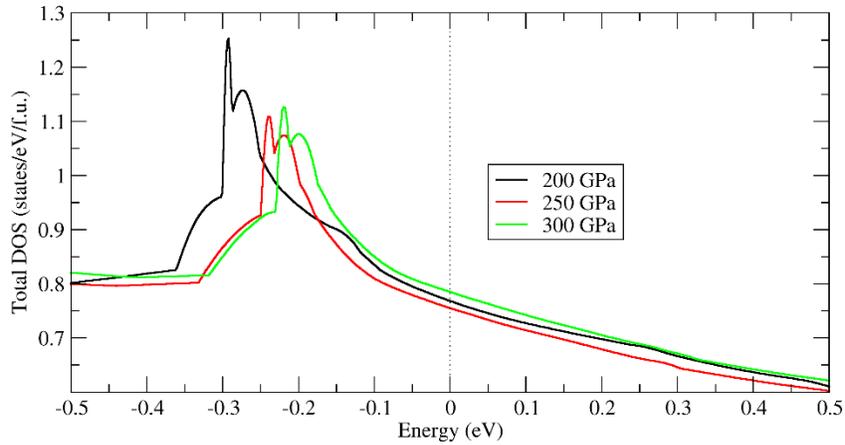

FIG. 5. (Color online) The total DOS of $Fm\bar{3}m$ LaH$_{10}$ at 200, 250 and 200 GPa.

Note that the peak in DOS is located not near the $\Gamma$ point, but at finite crystal momentum. This can be viewed as the presence of small pockets (Fig. 6) and can be treated similarly to the case in sulfur hydrides [31]. More specifically, the contributions to $T_c$ of a large Fermi surface and small pockets can be described as a two-gap scenario (see the review in [47]) with an additional small pairing gap created by the presence of the pockets. Then the additional small change in $T_c$ caused by the pockets is $\Delta T_c \propto \nu \propto p_{\text{F;poc}}$ (see [31]), where $\nu$ is the DOS on the pockets and $p_{\text{F;poc}}$ is the momentum for the pocket states. The energy scale corresponding to the pockets is small relative to the energy of the optical phonons, but this smallness does not violate the "Migdal criterion" thanks to the small value of corresponding coupling constant. An increase in pressure leads to a decrease in $p_{\text{F;poc}}$ and, correspondingly, to a decrease in $\Delta T_c$. The presence of the second gap and its evolution with pressure should be observable by tunneling spectroscopy.

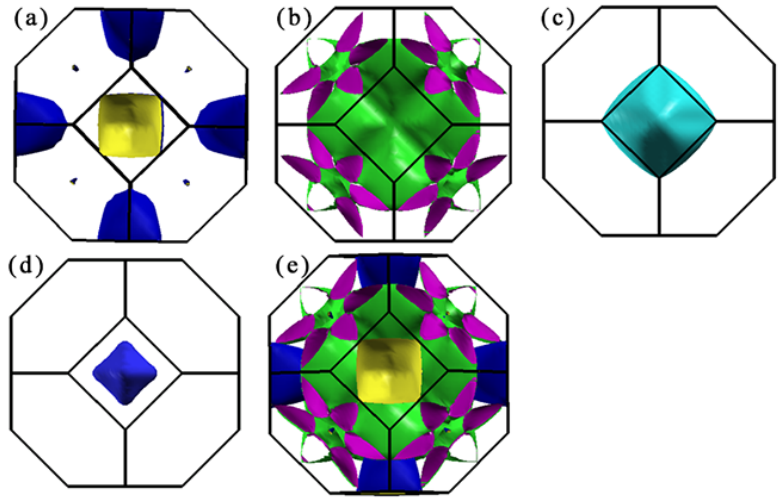

FIG. 6. (Color online) The Fermi surface of $Fm\bar{3}m$ LaH$_{10}$ calculated at 200 GPa. (a)-(d) The Fermi surface cross the first, second, third and fourth band. (e) The Fermi surface including all cutting bands.

The transition into the $Fm\bar{3}m$ state is similar to that of the transition into $Im\bar{3}m$, which, as discussed in [31,51] is a first-order transition. This transition is similar to that considered in [52] and it is caused by the electron-phonon interaction. It is tempting to relate the drastic increase in $T_c$ up to 250 K to the contribution of these pockets, and such an assumption was made in [53]. However, for the $H_3S$ compound this assumption is contradicted by the observation of a large isotope effect [2], see below. In addition, it is important to point out that a treatment based on this assumption and on Eq. (1) is not rigorous because it violates the "Migdal criterion" [54] (a similar argument can be found in [31]).

The evaluation of $T_c$ presented above is based on a picture of strong electron-phonon coupling provided by wide band supplemented by an additional weak interaction ($\lambda_{\mathrm{poc}} \ll 1$) with the pockets. This treatment is rigorous, self-consistent, and is in a good agreement with the experimental data.

**D. Isotope effect**

The isotope coefficient for $D \to H$ substitution in hydrides is very sensitive to the interplay of the optical and acoustic phonon modes. The hydrides of interest form a two-component lattice, which contains both heavy and light ions (in our case $M_H \ll M_{La}$). As a result, the optical modes correspond to the vibration of the hydrogen ions, whereas the acoustic modes reflect, mainly, the motion of heavy ions (see, e.g., [47]). As a result, $D \to H$ substitution greatly affects the value of $\widetilde{\Omega}_{\mathrm{opt}}$. Therefore, the value of the isotope coefficient correlates with the relative contributions of the optical and acoustic modes and, correspondingly, with $T_c$ (see, e.g., [55]). If the optical modes' contribution dominates (as is the case for $LaH_{10}$), one can expect that the isotope coefficient $\alpha$ will be large.

The isotope coefficient can be evaluated from the equation

$$\alpha = 0.5\left[1 - \frac{\lambda_{ac}}{\lambda_{opt}} \cdot \frac{\rho^2}{(\rho^2+1)^2}\right] \qquad (11)$$

derived in [31]. Expression (11) follows directly from Eq. (9) and the relation [31]

$$\alpha = 0.5\left(\frac{\widetilde{\Omega}_{opt}}{T_c}\right)(\partial T_c/\partial \widetilde{\Omega}_{opt}) \qquad (11')$$

The quantity $\rho$ is defined as in Eq. (9); Eq. (11') is valid in the harmonic approximation.

The values of the parameters entering Eq. (11) for various pressures are presented in Table I (for $\mu^* = 0.13$). Table II lists the corresponding values of the isotope coefficient.

Table II. The isotope coefficient of the $Fm\bar{3}m$ phase in $LaH_{10}$ at different pressures.

| $P$(GPa) | $\alpha$ |
|---|---|
| 200 | 0.45 |
| 250 | 0.47 |
| 300 | 0.46 |

It is seen that , indeed, the isotope coefficients are large and are close to the optimum value ($\alpha_{max} = 0.5$ in the harmonic approximation). Such large values (even larger $\alpha$ correspond to $\mu^* = 0.1$) reflect the dominant contribution of the optical modes for $LaH_{10}$. The measurements of the isotope coefficient for $LaH_{10}$ are described in [2]; the measured values are close to $\alpha = 0.45$.

## IV. DISCUSSION

The analysis presented above, provides a totally self-consistent description of the properties of a new high $T_c$ superconductor $LaH_{10}$. The material has some peculiarities, but as a whole the picture appears to be similar to that for the $H_3S$ compound, so that one can formulate the main factors leading to such high values of $T_c$ in the hydrides. The high value of the critical temperature is provided by remarkable combination of high frequency phonons (optical modes caused by the presence of light hydrogen ions) and the strong interaction of electrons, whose states belong to wide band, with these high frequency modes. A large strength of the interaction is related to the large number of hydrogen ligands in the unit cell. This is the case for the *Im$\bar{3}$m* structure of $H_3S$ compound, and the same is true for the *Fm$\bar{3}$m* structure of lanthanum hydride.

One can expect high value of $T_c$ for such hydrides as $YH_9$, $CaH_6$, etc. As was noted above, the structural transition of the $LaH_{10}$ from $R\bar{3}m$ or $C2/m$ into *Fm$\bar{3}$m* phase is similar that for the transition of $H_3S$ from *R3m* to *Im$\bar{3}$m* structure.

We did not study the impact of anharmonicity of the phonon spectrum. This problem was discussed in [23] and it was based on the approach described in [24-26], see also the review [29], Sec.3.1.4. However, one can raise a serious criticism of this approach. Indeed, it is based on the Born-Oppenheimer (BO) adiabatic approximation which allows us to separate the electronic and ionic motions, so that the total wave function can be written as a product of electronic and ionic wave functions. According to BO, the ionic motion is described by Schrödinger equation with the electronic energy term $\varepsilon(\mathbf{R})$ as the potential energy. However, as is known, the BO solution is not exact. It is valid if $(a/L) \ll 1$, $a$ is the vibrational amplitude, and $L$ is the lattice period (see, e.g., the review [47], Sec.2.2). In addition, the potential $\varepsilon(\mathbf{R})$ can be expanded near the equilibrium in a series of $(a/L)$ and the harmonic approximation corresponds to the second term in this expansion. The authors [24-26] are considering large anharmonicity, that is, the case, when the condition $(a/L) \ll 1$ is violated, but they are still using the BO approximation and the corresponding equation for ionic motion. However, such an approach is not self-consistent, since the electronic and ionic motions become inseparable (polaronic effects) and this equation is not valid anymore. According to the authors [24], a strong anharmonicity leads to a decrease in $T_c$. However, it is known that the double-well potential is the typical example of strong anharmonicity. In such a case, because of the dynamic Jahn-Teller effect we are dealing with an increase in the value of $T_c$ (see, e.g.,[56], review [47], Sec.3.9 ). Therefore, a strong anharmonicity should be treated with a considerable care. As mentioned above, we did not study the impact of anharmonicity. The corresponding corrections will be calculated elsewhere; one can expect that these corrections are not large, since the calculations of $T_c$ and the isotope coefficient performed, e.g., in [31,35] for $H_3S$ and for $LaH_{10}$ compound (see above) are in rather good agreement with the experimental data.

## V. CONCLUSIONS

In summary, we are focusing on the properties of the $Fm\bar{3}m$ phase of the lanthanum hydride. This phase is unique, since the compound displays the highest observed value of the critical temperature. The analysis described above allows us to evaluate the values of the critical temperature which can reach 254 K at 200 GPa. Moreover, its $T_c$ decreases with increasing pressure. We calculated also the value of the isotope coefficient which is about 0.45. The study is in a rather good agreement with the experimental data.

**ACKNOWLEDGEMENTS**

This work was supported by the National Natural Science Foundation of China (Nos. 11674122 and 51632002).